\documentclass[]{aa}

\usepackage{graphicx}
\usepackage{array}
\usepackage{tabularx}
\usepackage{txfonts}
\usepackage{hyperref}

\def\ltsima{$\; \buildrel < \over \sim \;$}
\def\simlt{\lower.5ex\hbox{\ltsima}}
\def\gtsima{$\; \buildrel > \over \sim \;$}
\def\simgt{\lower.5ex\hbox{\gtsima}}

\def\msun{{\rm M_{\odot}}}

\def\be{\begin{equation}}
\def\ee{\end{equation}}

\def\del#1{{}}
\def\ltsima{$\; \buildrel < \over \sim \;$}
\def\simlt{\lower.5ex\hbox{\ltsima}}
\def\gtsima{$\; \buildrel > \over \sim \;$}
\def\simgt{\lower.5ex\hbox{\gtsima}}

\def\Pon{P\=oniu\=a'ena }

\usepackage{xcolor}

\newcolumntype{L}{>{$}l<{$}}
\newcolumntype{C}{>{$}c<{$}}
\newcolumntype{R}{>{$}r<{$}}

\usepackage{booktabs}
\usepackage{siunitx}
\usepackage{comment}
\usepackage{longtable}

\usepackage{capt-of}

\begin{document}

   \title{Chaotic accretion can (still) explain early supermassive black hole growth}
   \titlerunning{Chaotic early SMBH growth}

   \author{K. Zubovas\orcid{0000-0002-9656-6281}\inst{1,2} \and M. Tartėnas\orcid{0009-0006-7373-180X}\inst{1}}

   \institute{Center for Physical Sciences and Technology, Saulėtekio al. 3, Vilnius LT-10257, Lithuania\\
              \email{kastytis.zubovas@ftmc.lt}
            \and
             Astronomical Observatory, Vilnius University, Saulėtekio al. 3, Vilnius LT-10257, Lithuania\\ 
             }

   \date{Received ...; accepted ...}

  \abstract
   {The origin of supermassive black holes (SMBHs) is still debated, with hypotheses ranging from massive seeds to super-Eddington growth to chaotic accretion. Recent discoveries of very high redshift SMBHs constrain their possible origin and growth scenarios.}
   {We investigate whether the chaotic accretion scenario can explain the most extreme high-redshift SMBH masses.}
   {We predicted the exponential growth timescale of black holes accreting chaotically in a slowly-spinning dark matter halo, in which uncorrelated gas inflows maintain a low spin and hence a low radiative efficiency. We compared this prediction against the growth timescales and seed masses implied by a compilation of all known SMBH mass estimates at $z>7$ and against the feeding and feedback conditions expected in such haloes.}
   {We found that in the period $7.6 < z < 10.6$, the most massive SMBHs likely grew very efficiently, with an exponential growth timescale $\tau \sim 32$~Myr, which is consistent with chaotic accretion predictions provided that growth was almost continuous. All high-redshift SMBHs with virial mass estimates could have grown from seeds with $M_{\rm seed} \lesssim 100 \, \msun$. We describe the evolution of the mass reservoir feeding such a black hole, suggesting that wind-driven feedback eventually removes most of the gas and changes the growth mode of black holes at $z \sim 7.5$. We make several predictions regarding the spins, host-galaxy correlations, and observable counterparts of high-redshift SMBHs that distinguish our model from its alternatives. These predictions will be testable in the near future.}
   {Chaotic accretion from stellar-mass seeds remains a viable origin for the earliest known SMBHs, without invoking massive seeds or super-Eddington growth.}


   \maketitle

\section{Introduction}

Supermassive black holes (SMBHs) exist in the centres of all large galaxies \citep{Kormendy2013ARAA}, but their origin and growth at early cosmic times still poses a conundrum \citep{Inayoshi2020ARAA, Chen2025MPLA}. The existence of SMBHs with masses $M_{\rm BH} > 10^9 \, \msun$ at a redshift $z > 7.5$ \citep{Yang2020ApJ, Wang2021ApJ} is difficult to explain within the `standard' luminous thin-disc accretion paradigm starting from stellar-mass seeds \citep{Inayoshi2020ARAA}, prompting the development of numerous alternative models of their assembly. More recent discoveries have pushed the record of most distant known SMBHs to $z > 14$ \citep{Naidu2026OJAp, Fabian2026MNRAS}, with virial mass estimates reaching $z > 10$ \citep{Maiolino2024AA} and at least seven such estimates at $z > 8$. Although their masses are lower, the shorter timeframe for growth means they may put an even stronger constraint on SMBH origins.

Any model of the assembly of the first SMBHs must align with further observational constraints on the properties of active galaxies at high redshift. For example, quasars at $z \sim 6-7$ have somewhat higher typical Eddington ratios than those in the Local Universe, but still generally accrete below the Eddington limit \citep{Onoue2019ApJ, Trefoloni2025arXiv}. Their duty cycles, i.e. the fraction of time spent in an active state, are also generally low, $\delta_{\rm AGN} < 0.02$ \citep{Arita2023ApJ, Eilers2024ApJ}; although see \citet{Roberts2026MNRAS} for an alternative, much higher value $\delta_{\rm AGN} > 0.1$, and numerical hydrodynamical simulations predict $\delta_{\rm AGN} \sim 1$ at $z > 4$ \citep{Habouzit2022MNRAS}.

Conversely, some recent discoveries suggest that SMBHs at very high redshift grew efficiently. A new sub-population of galaxies, known as Little Red Dots (LRDs), has been discovered spanning a wide redshift range, with greatest numbers between $4 < z < 8$ \citep{Matthee2024ApJ, Kocevski2025ApJ}. LRDs have been detected as far as $z > 10$ \citep{Tanaka2025ApJ}, suggesting that their fraction among all galaxies increases with redshift. Although they may represent a heterogeneous population \citep{Perez2026arXiv}, a large fraction of LRDs are most likely powered by AGN \citep{Kocevski2025ApJ, Li2025ApJ, Rusakov2026Natur} accreting at close to the Eddington limit \citep{Rusakov2026Natur, Tortosa2026AA}. Their compactness and associated rapid accretion can be explained by efficient gas transport within slowly-spinning dark matter haloes \citep{Pacucci2025ApJ}, which facilitates black hole growth by removing the angular momentum barrier. In addition, quasars at $z > 7$ exhibit either very low radiative efficiency $\epsilon < 0.01$ or very high obscuration, or both \citep{Davies2019ApJ}; both properties are conducive to rapid SMBH growth.

The hypotheses of SMBH origin and growth can be broadly divided into several groups. Massive seed hypotheses suggest that conditions at $z \sim 20-30$ allowed the formation of black holes of $10^4-10^5 \,\msun$ by direct collapse of gaseous halos \citep{Begelman2006MNRAS, Dijkstra2014MNRAS}, an intermediate quasi-star stage \citep{Volonteri2010MNRAS, Coughlin2024ApJ} or runaway collisions of stars and stellar remnants in clusters \citep{PortegiesZwart2004Natur, Alister2020MNRAS}. These objects can then grow via Eddington-limited accretion with a radiative efficiency of $\epsilon \sim 0.1$ to reach the observed masses at $z = 7-10$. A second group posits that primordial black holes, formed within the first few seconds after the Big Bang, could have reached the observed masses even with a lower initial mass due to having more time to grow \citep{Ziparo2025JCAP}. Finally, stellar-mass seeds forming around $z=25-30$ can grow to the observed high masses if the average radiative efficiency of accretion is low, either because of hyper-Eddington accretion \citep[e.g.,][]{Inayoshi2016MNRAS} or due to very low spin of the SMBH maintained by chaotic accretion with uncorrelated angular momentum directions \citep{King2008MNRAS}. 

Investigating the latter scenario, \cite{Zubovas2021MNRAS} showed that \Pon \citep{Yang2020ApJ} and J0313-1806 \citep{Wang2021ApJ}, the two SMBHs that provided the most stringent constraints at the time, could have grown from stellar-mass seeds provided that they accreted essentially continuously and chaotically, leading to very low spins and hence low radiative efficiencies, without requiring super-Eddington accretion rates. In this paper, we revisit the chaotic accretion scenario and show that it can explain the existence of currently known $z>7.5$ SMBHs with virial mass estimates. In particular, the most extreme black holes - i.e. the ones providing the strictest constraints on the seed mass and growth efficiency - could all have grown from seeds as small as $M_{\rm seed} \lesssim 100 \, \msun$, forming at $z \sim 30$. We also show that the chaotic accretion model, coupled with AGN feedback, naturally predicts the masses of SMBHs powering LRDs and that these masses should follow a modified, steeper $M-\sigma$ relation. Finally, we predict that the local $M-\sigma$ relation is established already at $z \sim 7.5$, which should also be the time at which the black-hole-to-stellar mass ratio peaks, and that most very-high-redshift SMBHs should be spinning slowly. All of these predictions will become testable in the near future. We also present a comprehensive list of SMBH mass estimates at $z > 7.5$.

The paper is structured as follows. We begin by reviewing the chaotic accretion model in Section \ref{sec:model} and specifying its predictions for black hole growth. In Section \ref{sec:growth}, we present the observed high-redshift SMBH sample and the implied growth timescales, as well as constraints on SMBH seed masses, showing their consistency with the chaotic accretion predictions. In Section \ref{sec:discussion}, we consider the implications of such growth for the observable properties of SMBHs at very high redshift (section \ref{sec:scenario}), the transition to feedback-regulated growth at $z\sim7.5$ and establishment of the SMBH-galaxy correlations (section \ref{sec:correlations}), the possibilities of distinguishing among the origin and growth scenarios (section \ref{sec:super_edd}) and the possible falsification of our model with more extensive discoveries of extreme black holes (section \ref{sec:completeness}). Finally, we summarise and conclude in Section \ref{sec:sum}.

All conversions between redshift and the age of the Universe assume the Planck 2018 cosmology \citep{Planck2020AA}; calculations are performed using the Ned Wright cosmology calculator \citep{Wright2006PASP}.

\section{SMBH growth by chaotic accretion}\label{sec:model}

Let us consider a slowly spinning dark matter halo, where the first stars form around $z \sim 30$ \citep{Johnson2008MNRAS}, evolve rapidly and create stellar-mass black holes with $M_{\rm seed} \sim 100 \, \msun$. The slow spin means that infalling intergalactic gas reaches the central regions at a significant fraction of the dynamical rate $\dot{M}_{\rm dyn} = f_{\rm g} \sigma^3 / G \approx 300 f \sigma_{200}^3 \, \msun$~yr$^{-1}$, where $f \equiv f_{\rm g}/0.16$ is the ratio of the gas fraction to the cosmological average. Rapid feeding is facilitated by mergers, which are expected to be almost constant for massive halos in this redshift range \citep{Tanaka2014CQGra}. The feeding rate is super-Eddington for most black holes:
\begin{equation}
    \frac{\dot{M}_{\rm dyn}}{\dot{M}_{\rm Edd} \left(M_{\rm BH}\right)} = \frac{f_{\rm g} \kappa \epsilon \sigma^3 c}{4 \pi G^2 M_{\rm BH}} = \frac{M_\sigma}{M_{\rm BH}} \frac{\epsilon c}{4 \sigma} \approx 38 \frac{M_\sigma}{M_{\rm BH}} \frac{\epsilon}{0.1} \sigma_{200}^{-1}.
\end{equation}
Here, $M_\sigma = f_{\rm g}\kappa \sigma^4 / \left(\pi G^2\right) \approx 3\times10^8 \sigma_{200}^4 \, \msun$ is the SMBH mass given by the theoretical $M-\sigma$ relation from \citet{King2010MNRASa}. Since $M_\sigma \geq M_{\rm BH}$, accretion at only a small fraction of $\dot{M}_{\rm dyn}$ is enough to exceed the Eddington rate, potentially by many orders of magnitude. Turbulence, shocks and star formation can significantly reduce this rate; however, galaxy-scale hydrodynamical simulations investigating the feeding of central black holes generally find that sustained super-Eddington feeding is possible at rates $\dot{m} \equiv \dot{M}/\dot{M}_{\rm Edd} \sim 1-100$ \citep{Massoneau2023AA, Lupi2024AAb, Mehta2026NatAs, Shi2026ApJ}. This is generally below the threshold $\dot{m}_{\rm crit} \sim 40-500$ required for the accretion flow to trap photons and flow efficiently into the black hole \citep[][; see also Section \ref{sec:super_edd}]{Inayoshi2016MNRAS, Johnson2022ApJ}. As a result, the sustained large-scale super-Eddington feeding of the nucleus is unlikely to lead to equivalently super-Eddington growth of the black hole. Instead, a solution sometimes called hyperaccretion \citep{King2008NewAR} emerges: the accretion flow becomes quasi-spherical, and radiation and mechanical outflows push the excess gas out, maintaining an approximately Eddington-limited gas inflow. This occurs at radii interior to $R_{\rm sph} \approx \dot{m} R_{\rm s} \approx 10^{-5}-10^{-3} \left(M_{\rm BH}/10^8 \, \msun\right)$~pc, where $R_{\rm s}$ is the Schwarzschild radius.

Individual accretion episodes likely increase the black hole mass only by a small amount, because the accretion disc fragments into self-gravitating clumps beyond a radius containing a mass $M_{\rm d} \sim \left(H/R\right) M_{\rm BH} \sim 10^{-3} M_{\rm BH}$, where $H/R$ is the disc aspect ratio \citep{Pringle1981ARAA, King2008MNRAS}. The nearly continuous infall of gas from larger scales leads to the black hole growing through a very large number of such events. The chaotic geometry of feeding streams leads to individual episodes having uncorrelated directions of initial angular momentum vectors. The Lense-Thirring effect causes the disc to precess; eventually, the central parts of the disc align or counter-align with the SMBH spin vector. Counter-alignment occurs if the initial angle $\theta$ between the black hole spin angular momentum $\vec{J}_{\rm BH}$ and the orbital angular momentum of the central warped part of the disc $\vec{J}_{\rm d}$ satisfies $\cos \theta < - |\vec{J}_{\rm d}|/2 |\vec{J}_{\rm BH}|$ \citep{King2005MNRAS}. The disc material then accretes on to the black hole, adding its angular momentum at the innermost stable circular orbit (ISCO), $J_{\rm ISCO} \equiv M_{\rm d} j_{\rm ISCO}$, to that of the black hole. In the case of counter-alignment, the black hole's angular momentum is reduced. This happens in fewer than half of accretion episodes, but a counter-aligned disc has a much higher ISCO radius, so the individual spin-down is stronger than the spin-up in an average aligned episode. Over many episodes, the SMBH dimensionless spin parameter $a_{\rm BH}$ tends to oscillate but decreases slowly; it is essentially always $a_{\rm BH} < 0.7$, which corresponds to a radiative efficiency $\epsilon < 0.1$ \citep{Bardeen1972ApJ}, and drops to $a_{\rm BH} < 0.4$ once $M_{\rm BH} \gtrsim 10^{4-5} \, \msun$ \citep[the higher value corresponds to a model where 20\% of accretion events are aligned; ][]{King2008MNRAS, Zubovas2021MNRAS}. The spin parameter range typical for massive black holes, $0.2 < a_{\rm BH} < 0.4$, results in $0.064 \lesssim \epsilon \lesssim 0.075$. Radiative efficiency limits the black hole growth rate through its connection to the accretion luminosity; the resulting black hole mass evolution with time is
\begin{equation}\label{eq:mbh-time}
    M_{\rm BH}\left(t\right) = M_{\rm BH,0} \times \exp \left(\frac{t-t_0}{\tau}\right),
\end{equation}
where
\begin{equation} \label{eq:salpeter}
    \tau = \frac{\kappa_{\rm e.s.} c}{4\pi G} \frac{\epsilon}{1-\epsilon} \frac{1}{\delta \left\langle l_{\rm E} \right\rangle} \approx 43.7 \left(\frac{\epsilon}{0.1}\right) \left(\frac{0.9}{1-\epsilon}\right) \frac{1}{\delta \left\langle l_{\rm E} \right\rangle} \, {\rm Myr}.
\end{equation}
In these two equations, $M_{\rm BH,0}$ is the black hole seed mass, which forms at time $t_0$, $\kappa_{\rm e.s.} = 0.34$~cm$^2$~g$^{-1}$ is the electron scattering opacity, $c$ is the speed of light, $G$ is the gravitational constant, $\delta$ is the average AGN duty cycle and $\left\langle l_{\rm E} \right\rangle$ is the average Eddington ratio during each AGN episode. For $0.064 < \epsilon < 0.075$, the growth timescale falls in the range $26.9$~Myr$< \tau < 31.9$~Myr. In the next section, we show that such growth timescales can explain the presence of even the most extreme currently measured SMBH masses.

\section{High-redshift SMBHs}

\subsection{Observed masses and possible growth histories} \label{sec:growth}

\begin{figure}
\includegraphics[width=0.49\textwidth, trim={0 0 0 0}, clip]{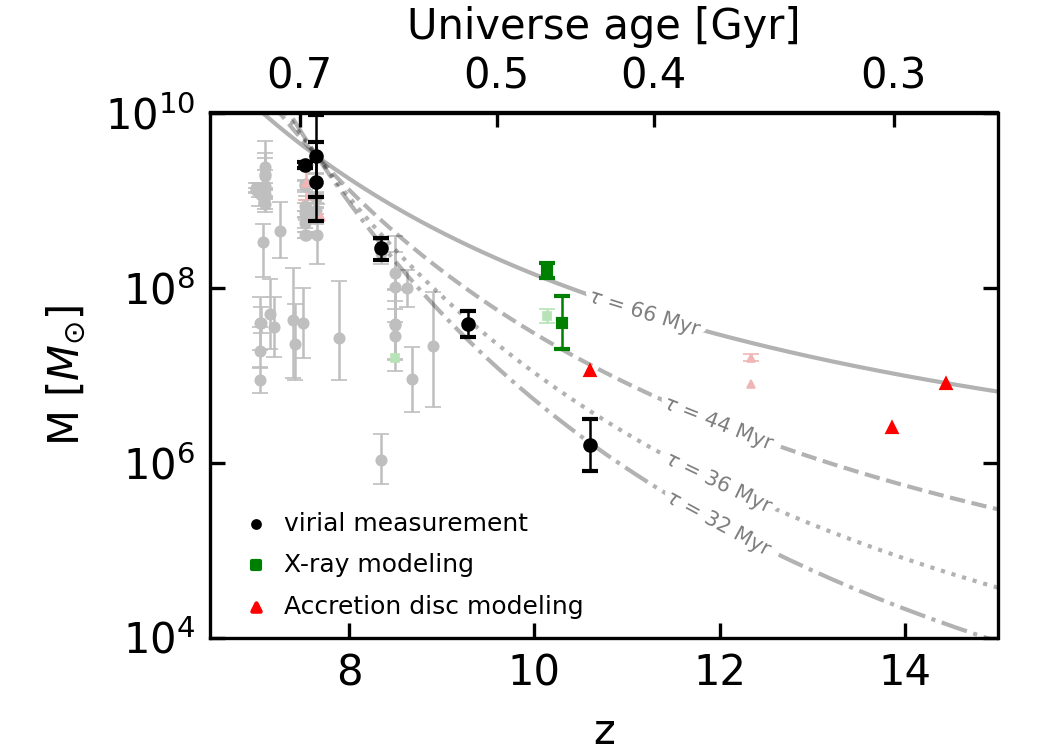}
\caption{High-redshift SMBH mass estimates. The black/grey points show virial (usually MgII or $H\beta$ line) mass estimates, green points show mass estimates from X-ray luminosity and spectrum, and red points show mass estimates using UV spectral slopes assuming Eddington-limited thin-disc accretion. Darker symbols correspond to the SMBHs for which we plot seed mass constraints in Figure \ref{fig:seed_constraint}. We only show the highest mass estimate for each object, except for J0313-1806, for which we show two; the full list is given in Table \ref{tab:SMBH_data}. The grey lines indicate four possible growth histories of the higher (Mg II-based) J0313-1806 mass estimate with different average growth timescales $\tau$, as indicated in the figure.}
\label{fig:growth_histories}
\end{figure}

We compiled a list of SMBH mass estimates at $z>7.5$, plus several particularly massive SMBHs at $7 < z < 7.5$. The full table is given in the Appendix, and we show the data in Figure \ref{fig:growth_histories}. We chose to distinguish between mass estimates made using different methods, mainly because virial estimates are the most robust (see Appendix), while X-ray and UV-based ones are systematically higher at a given redshift, so may be overestimates.

The grey lines show several possible SMBH growth histories (eq. \ref{eq:mbh-time}), assuming continuous growth with a timescale $\tau$ as given in the Figure. They are anchored at the highest, Mg II-based, mass estimate of J0313-1806, $M_{\rm J03} = 3.1\times10^9\,\msun$ \citep{WolfBanados2026}, at $z=7.64$. Assuming that the most massive known black holes at each redshift constitute a true upper boundary of their mass distribution (we discuss the validity of this assumption in Section \ref{sec:completeness} below), we see that J0313-1806 could have grown from an SMBH equivalent to CAPERS\_LRD\_z9 at $z=9.3$ with a growth timescale $\tau \approx 36$~Myr (grey dotted line) and from GN-z11 at $z = 10.6$ with $\tau \approx 32$~Myr (dash-dotted line). Under the extreme case of continuous and Eddington-limited accretion, i.e. $\delta = \langle l_{\rm E} \rangle = 1$, these timescales could be achieved if the AGN had a radiative efficiency of $\epsilon \approx 0.075-0.082$ (see eq. \ref{eq:salpeter}). Comparing these values with the predictions from the previous section, we see that growth via chaotic accretion is possible as long as the product $\delta \langle l_{\rm E} \rangle \gtrsim 0.85$. Growth from black holes with masses estimated using X-ray or UV spectral properties could have been achieved with longer growth timescales up to $\tau \approx 66$~Myr (solid line); growth at $\epsilon = 0.1$, corresponding to $\tau \approx 44$~Myr, is shown with a dashed line for reference.

\subsection{Possible seed masses} \label{sec:seed_constraints}

\begin{figure}
\includegraphics[width=1.\columnwidth,trim={0 0 3cm 0},clip]{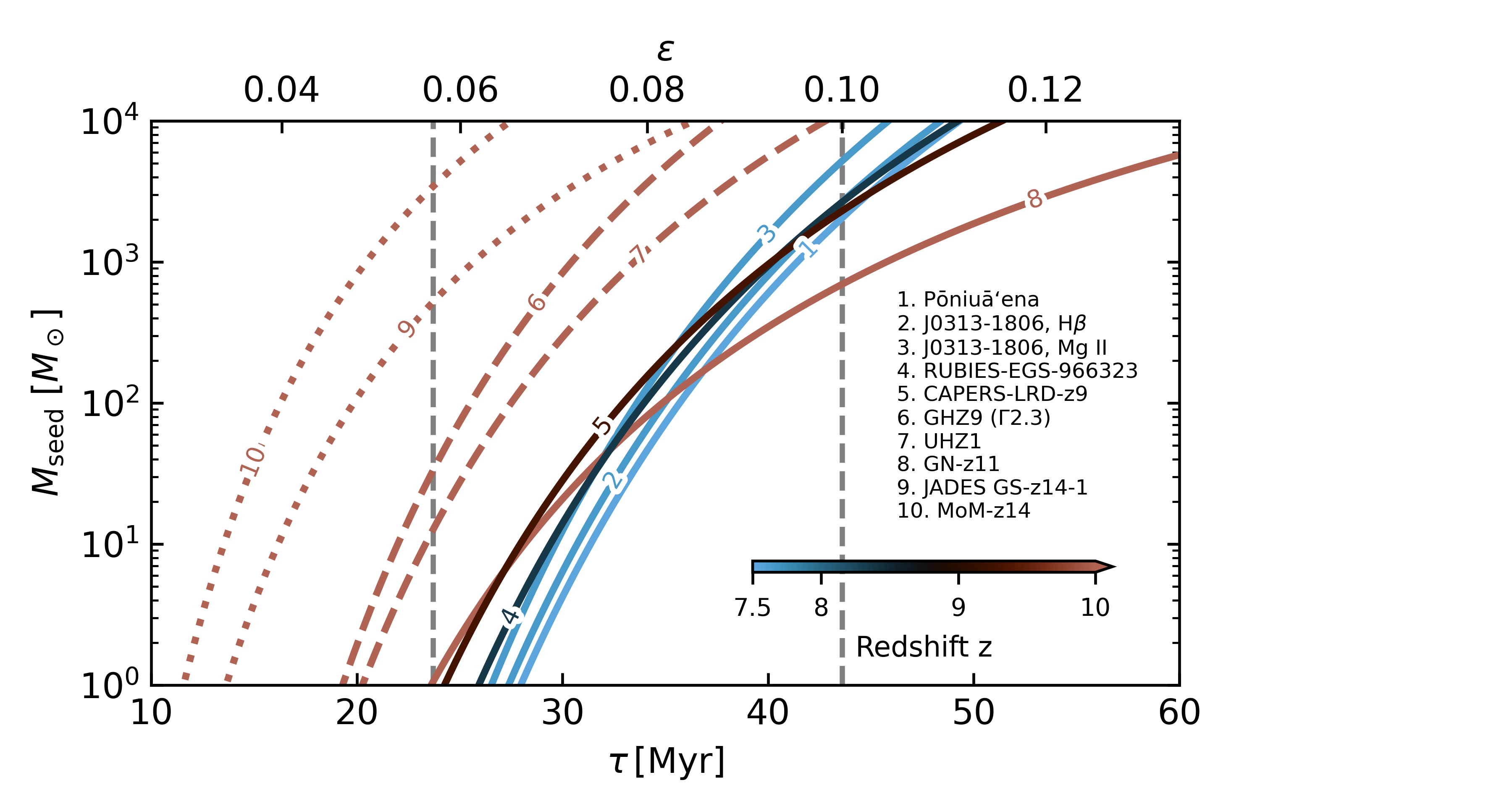}
\caption{Relation between the average growth timescale and required seed mass for selected SMBHs with $z > 7.5$ from Table \ref{tab:SMBH_data}, assuming seed formation at $t = 100$~Myr ($z = 30$). Lines are labelled in order of increasing redshift at observation, with the list of objects given on the right. Solid, dashed and dotted curves represent SMBHs with virial, X-ray-based and UV-based mass estimates, respectively. Line colours show the observed SMBH redshift, with the colour scale on the bottom right. The top axis shows values of the radiative efficiency $\epsilon$ corresponding to the growth timescales, assuming $\delta = \left\langle l_{\rm E}\right\rangle = 1$. Vertical dashed lines highlight $\epsilon = 0.1$, typically assumed in SMBH growth calculations, and $\epsilon = 0.057$, corresponding to a non-rotating black hole.}
\label{fig:seed_constraint}
\end{figure}

We can use equation \ref{eq:mbh-time} to connect the observed SMBH masses, their potential seed masses $M_{\rm seed}$ at an assumed formation time $t_{\rm form}$ and the growth timescale. We plot these relations, assuming a formation redshift $z_{\rm form} = 30$, corresponding to $t_{\rm form} = 100$~Myr \citep{Johnson2008MNRAS}, in Figure \ref{fig:seed_constraint} for selected SMBHs observed at $z > 7.5$; we choose the objects with virial mass estimates that produce the strongest constraints, and two each of the estimates based on X-ray and on accretion disc spectral models for illustration. SMBHs with virial mass estimates allow smaller $M_{\rm seed}$ values than those from the other two subsamples; the growth timescale that connects the extreme virially-estimated SMBH masses, $\tau = 32$~Myr, allows the growth of all virially-estimated SMBHs from seeds $M_{\rm seed} \leq 100 \, \msun$, consistent with predictions of first star remnant masses \citep{Hirano2014ApJ, Stacy2016MNRAS}, without requiring super-Eddington accretion episodes. It is worth noting that approximately the same initial seed mass and growth timescale can explain the six virial SMBH mass estimates. This strengthens the validity of our assumption that they constitute a true upper bound on SMBH masses at those redshifts and that SMBH growth during the period $7.6 < z < 10.6$ was approximately continuous.  X-ray- and UV-based mass estimates suggest seed masses $M_{\rm seed} > 500 \, \msun$, even if the growth timescale is similarly short.

Varying the formation redshift has a relatively small effect on the required seed masses. If black holes started forming at $z = 40$ ($t = 66$~Myr), the required $M_{\rm seed}$ is approximately a factor of three lower. If they formed only at $z = 20$ ($t = 180$~Myr), the required $M_{\rm seed}$ increases by a factor of 10. In the latter case, formation of the observed black holes from stellar mass seeds is only possible if the average growth timescale is $\tau \lesssim 27$~Myr, which is the lowest possible under the chaotic accretion model \citep{Zubovas2021MNRAS}.

\section{Discussion} \label{sec:discussion}

At early times after seeding, the black hole is too small to have a noticeable effect on the infalling gas reservoir. We label this period the ``supply-limited growth phase''. Eventually, the SMBH grows enough to affect the surrounding material, primarily via wide-angle disc winds that can drive large-scale outflows \citep{King2015ARAA}; the main alternative feedback mechanism, jets, is not applicable because low black hole spin leads to them being weak or non-existent \citep[e.g.,][]{Curd2023MNRAS}. Eventually, feedback becomes strong enough to quench the SMBH growth, and the system transitions to a ``feedback-limited'' phase of evolution. In the next two subsections, we quantify this simple picture and show how it leads to testable predictions about LRDs and the black hole - galaxy correlations. We then discuss the potential for distinguishing between the SMBH growth mechanisms and describe how future discoveries of SMBHs at ever higher redshifts could test our model.

\subsection{Evolution of the feeding reservoir} \label{sec:scenario}

The luminosity emitted by the AGN during hyperaccretion scales logarithmically with the feeding rate, $L_{\rm AGN} = L_{\rm Edd} \left(1+\ln \dot{m}\right)$. Simulations show that $\dot{m} \lesssim 100$ \citep{Massoneau2023AA,Lupi2024AA,Mehta2026NatAs,Shi2026ApJ}, which results in $L_{\rm AGN} < 5 L_{\rm Edd}$. A strongly intermittent accretion flow may produce even lower luminosity $L_{\rm AGN} <L_{\rm Edd}$, while the black hole growth rate remains $\sim \dot{M}_{\rm Edd}$ \citep{Inayoshi2016MNRAS}.

In principle, the accreting black hole can power outflows at early times as well, but their velocity is much lower than that of the infalling gas. Within the wind-driven outflow paradigm, winds shock against the interstellar medium and transfer most of their energy rate to the surrounding gas; under normal conditions it is $\sim \epsilon L_{\rm AGN} / 2$, but can reach $\sim L_{\rm AGN}$ if the AGN is heavily obscured \citep{Zubovas2019MNRASb}. The velocity of the outer shock of this energy-driven outflow is (cf. \citealt{King2010MNRASa, Zubovas2023MNRAS})
\begin{equation}
    \dot{R}_{\rm out} \approx 850 \left(\frac{\epsilon}{0.1}M_8 \frac{l_{\rm E}}{f}\right)^{1/3} \sigma_{200}^{-2/3} \, {\rm km}\,{\rm s}^{-1},
\end{equation}
where we scale the black hole mass to $10^8 \, \msun$ and $\sigma$ to $200$~km~s$^{-1}$. Assuming $f = l_{\rm E} = 1$, we can derive a condition on the black hole mass necessary to have $\dot{R}_{\rm out} > \sigma$:
\begin{equation} \label{eq:mbh_for_outflows}
    M_{\rm BH} > 1.3\times10^6 \frac{0.1}{\epsilon} \sigma_{200}^5 \, \msun.
\end{equation}
Accounting for the possibility that $l_{\rm E} > 1$ can reduce this limit by a factor of a few; on the other hand, the accumulation of gas and baryon domination in the central part of the halo can increase the limit by a similar factor. This criterion shows that black holes can be fed without feedback producing any large-scale outflows until they reach masses of order $10^5-10^7 \, \msun$ in halos with central velocity dispersions $100 {\rm \ km \, s}^{-1} < \sigma < 300\ {\rm km \, s}^{-1}$.

Once the black hole mass exceeds the limit in eq. \ref{eq:mbh_for_outflows}, large-scale outflows can clear some sightlines and make the nucleus observable. However, this does not quench SMBH growth, because most of the outflow energy escapes through low-density channels, leaving the dense gas unaffected \citep{Zubovas2014MNRASb, Bourne2014MNRAS, Tartenas2026AA}. Significant modulation of the accretion rate only becomes possible when the SMBH mass reaches $M_\sigma$. At this point, the AGN wind momentum rate balances the weight of the infalling gas \citep{King2003ApJ, King2010MNRASa} and further growth is stalled.

So our model predicts a natural connection between LRDs and the brightest high-redshift quasars. They are members of the same population of rapidly-growing black holes, that first become visible once they reach masses comparable to those inferred for LRDs \citep{Rusakov2026Natur}, and cease rapid accretion once they reach the masses typical of quasars. The number density of LRDs has been variously estimated to range from $\Phi_{\rm LRD} \sim 10^{-5}$~Mpc$^{-3}$ \citep{Matthee2024ApJ} to $\Phi_{\rm LRD} \sim 10^{-4\pm0.1}$~Mpc$^{-3}$ \citep{Perez-Gonzalez2024ApJ}, with numbers generally declining at $z > 8$. Following the same calculation as in \citet{Zubovas2021MNRAS}, which gives the density of stellar black hole seeds $\Phi_{\rm seed} \sim 0.4$~Mpc$^{-3}$, we see that only a fraction $f_{\rm LRD} < 2.5\times10^{-4}$ of all seeds needs to encounter the conditions described in this scenario to explain the LRD population. Furthermore, only $f_{\rm LRD-q} \sim 10^{-3}-10^{-4}$ of LRDs need to continue growing under such extreme conditions to recreate the quasar space density at $z \sim 7.5$. This is consistent with the results of some semi-analytical models suggesting that extremely favourable conditions are necessary for SMBHs to grow to observed masses from stellar-mass seeds \citep{Griffin2019MNRAS, Trinca2026MNRAS, Jeon2025ApJ}. We note that the connection between black hole seeds and massive quasars is not contingent on them passing through the LRD phase; if it turns out the LRDs are inconsistent with the chaotic accretion scenario, this does not invalidate the scenario's applicability to the extreme SMBHs at $z > 7.5$.

The strong dependence of the critical mass for launching outflows on $\sigma$ (eq. \ref{eq:mbh_for_outflows}) suggests that we should find a strong correlation between LRD masses and their host galaxy velocity dispersions, especially at the lower end of the mass range. Our scenario naturally explains the LRD compactness and the possible importance of outflows in producing their broad lines \citep{Noboriguchi2023ApJ, Akins2025ApJ}. Furthermore, if the emitted luminosity can only escape through narrow gaps, it can become significantly beamed; this leads to the observed flux being much higher than isotropic \citep{King2009MNRAS, King2024MNRAS}. This effect can explain the spectra of AGN such as the one in MoM-z14 and others at $z > 10$, whose SMBH masses, when estimated assuming isotropic emission and Eddington luminosities, are much higher than predicted by our model \citep{Fabian2026MNRAS}, and whose luminosities, when compared with virially determined masses, are significantly super-Eddington \citep{Maiolino2024Natur}.

\subsection{Establishing the black hole - galaxy correlations} \label{sec:correlations}

We predict that galaxies as far as $z\sim7.5$ should have SMBH masses consistent with the local $M-\sigma$ relation. Currently, most analyses of the redshift evolution of this relation extend only to $z<2$. They generally suggest that the relation either doesn't change \citep{Salviander2013ApJ} or becomes flatter and its normalisation increases slightly \citep{Silverman2022ApJ} as redshift increases. JWST-based studies of high-redshift AGN-host galaxies show that at least three galaxies at $z \sim 6$ lie on the local $M-\sigma$ relation \citep{Onoue2025NatAs, Liu2026ApJ}, and in general SMBHs at $2 < z < 7$ lie much closer to the local $M-\sigma$ relation than to the local $M-M_*$ relation \citep{Maiolino2024AA, JuodzbalisMaiolino2026}, in agreement with our prediction (see below). As more stellar velocity dispersion measurements at higher redshift become available, they will serve as a powerful test of our model.

Once the SMBH reaches $M_\sigma$, any subsequent feeding events result in powerful outflows that disrupt the gas reservoirs. This is consistent with the relatively high incidence of outflows in $z \sim 6$ quasars \citep{Bischetti2022Natur, Bischetti2023ApJ} and their high power when compared with quasars at lower redshifts \citep{Liu2026Natur}. The outflows help maintain a relatively low duty cycle, consistent with observed quasar properties at redshifts below $z \sim 7.5$. The SMBH mainly grows via mergers and inefficient accretion in lockstep with its halo. Inefficient growth is highlighted by the typical observed $z\sim 6$ quasar parameters of $\delta = 0.02$ \citep{Arita2023ApJ} and $\left\langle l_{\rm E} \right\rangle = 0.33$ \citep{Onoue2019ApJ}, which, inserted into eq. \ref{eq:salpeter}, give $\tau \approx 6.6 \left(\epsilon/0.1\right)$~Gyr.

We also predict that the most massive black holes at $z\sim7.5$ should be somewhat overmassive compared to their galaxies. Our model does not make any predictions about the host galaxy star formation rate, but we can compare observed values with the predicted SMBH growth rates. The Eddington rate for a $10^9 \, \msun$ SMBH with a low radiative efficiency $\epsilon = 0.07$ is $\dot{M}_{\rm Edd} \approx 31 \, \msun$~yr$^{-1}$. This is only an order of magnitude below the star formation rates in $z \sim 7-8$ bright quasar host galaxies \citep{Venemans2017ApJ, Neeleman2019ApJ, Yang2020ApJ, Wang2021ApJ} and comparable to the average star formation rates in individual $z > 8$ galaxies \citep{Harikane2023ApJ}. If the black hole grows from low-mass seeds, either continuously at the Eddington rate, or through short super-Eddington bursts, and follows the tracks identified in Section \ref{sec:growth}, its mass and the Eddington rate increase by almost two orders of magnitude between $z \sim 7.5$ and $z \sim 9$, significantly more than the change in observed star formation rates. So the $M_{\rm BH}/M_*$ ratio should grow with cosmic time until the SMBH growth mode changes, and reach a peak at $z \sim 7.5$. Once the SMBH begins to regulate its own growth via feedback, the stellar population can grow relatively faster, and the ratio $M_{\rm BH}/M_*$ should decrease.

The observational evidence regarding the $M_{\rm BH}/M_*$ ratio at different redshifts is somewhat mixed. Several studies suggest that high-redshift SMBHs are overmassive with respect to the stellar masses of their host galaxies when compared with the local relation \citep{Pacucci2023ApJ, Maiolino2024AA}, in agreement with our prediction. However, this conclusion isn't universally accepted: \citet{Li2025ApJb}, \citep{Silverman2025ApJ} and \citet{Ziparo2026arXiv} claim this may be due to selection biases, while \citet{Noboriguchi2023ApJ} and \citet{King2024MNRAS} argue that the black hole masses may be overestimated. Furthermore, many other studies, exploring a range of somewhat lower redshifts $2 \lesssim z \lesssim 5.7$, find that many SMBHs lie close to the local $M_{\rm BH}/M_*$ relation \citep{Carnall2023Natur, Hoshi2025ApJ, Ren2025MNRAS, Li2025arXiv, Ito2026AA}. This can be explained if the reduction of the mass ratio occurs relatively quickly, within the first 1-3 Gyr of cosmic time. Larger and deeper samples will be required to track the true evolution of the $M_{\rm BH}/M_*$ ratio and provide another opportunity to discriminate among the origin scenarios.

\subsection{Distinguishing between growth and origin mechanisms} \label{sec:super_edd}

The two primary alternatives to the chaotic accretion scenario, massive (and primordial) seed models and super-Eddington accretion, make predictions that distinguish them from our model. A massive seed with $M_{\rm seed} = 10^4 \, \msun$ \citep[a rather conservative value; cf.][]{Inayoshi2020ARAA} forming at $z_{\rm form} = 20\ (30)$, corresponding to $t_{\rm form} = 180 \ (100)$~Myr, must grow with an average growth timescale of $\tau \approx 40 \ (46)$~Myr to reach the mass $M_{\rm J03}$ by $z = 7.642$. This is consistent with continuous growth at $\epsilon \sim 0.1$. However, such an SMBH would have reached a mass of $M\left(z=10.6\right) \approx 6.8 \ (15.1) \times 10^6 \ \msun$ by $z = 10.6$. Even the lower value exceeds the upper uncertainty limit of the virial mass estimate of GN-z11 \citep{Maiolino2024Natur}. Heavier seeds or earlier formation would only exacerbate this tension. Massive seeds can be reconciled with the data if the seed forms very late, at $z \lesssim 14$, or doesn't grow appreciably until this redshift. In either case, the subsequent growth to reach the mass of J0313-1806 must proceed on a short timescale $\tau \approx 32$~Myr. So massive seeds do not alleviate the need for rapid growth with low average radiative efficiency.

Super-Eddington accretion can also lead to the low required average growth efficiency $\epsilon \lesssim 0.075$. The physical viability of this scenario depends on whether the sustained large-scale feeding rate of the galactic nucleus exceeds the threshold required for efficient super-Eddington accretion rate on event horizon scales. The threshold rate is sensitive to many assumptions about the geometry of the system and the relevant physical processes. Spherical and quasi-spherical models generally estimate the rate to be as large as $\dot{m}_{\rm crit} \sim 500$ \citep{Inayoshi2016MNRAS, Takeo2020MNRAS}. Analytically, the threshold has been estimated as $\dot{m}_{\rm crit} = 2/\epsilon$ \citep{Johnson2022ApJ}, with low radiative efficiencies of a few per cent resulting in $\dot{m}_{\rm crit} \gtrsim 40$, bringing the estimates in line with the sustained accretion rates found in larger-scale simulations. On the other hand, if the SMBH is rapidly spinning, the energy released during accretion can escape along the jets, increasing radiative efficiency to $\epsilon > 0.1$ while simultaneously allowing the black hole to accrete most of the infalling material \citep{Curd2023MNRAS}. Such a scenario, however, is somewhat contradictory. Rapid black hole spin requires the gas streams feeding the black hole to have approximately coherent directions, which is achieved when the whole gas distribution has non-negligible spin. On the other hand, spin creates an angular momentum barrier that the gas must overcome to accrete. So, to feed the black hole efficiently while also spinning it up, the gas angular momentum must be low but nonzero and have a consistent direction. Such a situation may be difficult to maintain over long periods.

This produces a distinction between the chaotic accretion and super-Eddington growth scenarios: the former predicts that the most massive black holes should have very low spins, required to keep the radiative efficiency low, while the latter predicts high spins, powerful jet activity and high radiative efficiency. Current estimates of the black hole spin distribution favour large values \citep{Reynolds2021ARAA}, although with a weak trend of lower values at the highest masses. However, various observational effects can lead to overestimated spins \citep{Mummery2025MNRAS}. In addition, the current data only covers much lower redshifts, which, as we argued, are not representative of the early SMBH growth conditions. Upcoming observatories such as NewAthena \citep{Cruise2025NatAs} will be able to constrain the spin and accretion properties of these black holes \citep{Aird2013arXiv, Dovciak2013arXiv}.

\subsection{Completeness of the extreme redshift black hole sample} \label{sec:completeness}

When presenting the possible SMBH growth scenarios, we assumed that the upper envelope of the observed distribution is equivalent to that of the true distribution. This is an obvious simplification that depends on the completeness of the sample. Assuming a space density of bright quasars powered by SMBHs with $M \geq 10^9 \, \msun$ of $\Phi\left(M_{1450} < -26\right) = 0.39 \times 10^{-0.78\left(z-6.7\right)}$~Gpc$^{-3}$ \citep{Wang2019ApJ}, we would expect a space density of $0.038\ (0.0063)$~Gpc$^{-3}$ at $z = 8 \ (9)$. To find an average of one bright quasar, the survey would have to cover an area of $\sim 2400 \ (24400)$~deg$^2$ at these redshifts, assuming a $\Delta z = 1$ redshift bin. Present-day wide-field surveys have the necessary area coverage \citep{Wang2019ApJ}, but are not deep enough and/or cannot cover the required wavelength range to capture the Lyman breaks at such high redshift. The deep JWST surveys, such as UNCOVER \citep{Weaver2024ApJS}, CEERS \citep{Finkelstein2025ApJ} and JADES \citep{Eisenstein2026ApJS}, that yielded most of the $z > 8$ SMBH discoveries cover only a very small fraction of the sky, $A \ll 1$~deg$^2$. So the current discoveries cannot provide robust constraints on the high end of the black hole mass function at these extreme redshifts. Ongoing and upcoming surveys, such as Euclid \citep{Euclid2019AA}, Roman and Rubin \citep{Tee2023ApJ}, will probe deeper into the cosmic dawn and are already beginning to fill this gap \citep{Yang2026AA}. The most important prediction of our model is the existence of a transition in the slope of the upper envelope of the SMBH mass distribution. If upcoming observations confirm that the highest black hole masses begin to decrease rapidly at $z > 8$, it would strengthen our argument; conversely, if more very massive black holes at much higher redshifts are discovered, the massive-seed, slower-growth model of SMBH origin would become more viable. \citet{Zubovas2021MNRAS} predicted that finding a black hole with $M > 10^9 \, \msun$ at $z > 9.1$ would invalidate the chaotic accretion model even under the most generous assumptions, because there would not have been enough time for it to have grown from a stellar-mass seed even under continuous Eddington-rate accretion.

We also note that in our chaotic accretion model, the rapidly growing black holes are heavily obscured, making their discovery more difficult regardless of their mass or accretion rate. This further limits the constraining power of non-detections, especially below the predicted transition when large-scale outflows become viable (see eq. \ref{eq:mbh_for_outflows}). However, methods that are less sensitive to obscuration, such as observations of radio AGN or gravitational waves, could resolve the issue \citep{Pacucci2019BAAS}.

\section{Summary} \label{sec:sum}

We analysed the available data on $z > 7$ SMBH masses, assuming that the most massive detected objects represent the true upper end of the mass function at their redshifts. We showed that the population of SMBHs with virially-estimated masses is consistent with very efficient growth, with an average exponential growth timescale $\tau \sim 32$~Myr, at least between $z \sim 10.6$ and $z \sim 7.6$. Such growth scenarios are compatible with chaotic accretion, which requires neither massive seed black holes nor super-Eddington accretion episodes. Assuming that this growth timescale was maintained at earlier times suggests that even extreme high-redshift SMBHs could have grown from seeds with masses as low as $M_{\rm seed} \lesssim 100 \, \msun$ that formed at $t_{\rm form} = 100$~Myr after the Big Bang. The chaotic accretion scenario implies black hole feeding from an effectively unlimited gas reservoir maintained by strong gas inflows in a slowly spinning dark matter halo. The uncorrelated inflow directions maintain a low black hole spin, which leads to the low radiative efficiency $\epsilon \sim 0.064-0.075$ necessary for rapid growth. Initially, the reservoir keeps the system undetectable, until winds generated in the accretion flow are able to drive large-scale outflows. This occurs once the SMBH mass reaches $M \sim 10^5 - 10^7 \, \msun$, with a strong dependence on the velocity dispersion. The outflows clear some sightlines and the system becomes visible, possibly as a LRD. Efficient black hole growth continues until the SMBH mass reaches the $M-\sigma$ relation. At this point, momentum-driven outflows become powerful enough to clear most of the gas reservoir and shut down further growth. At later times, efficient feedback prevents rapid growth, so the maximum SMBH masses at lower redshifts are not much higher than at $z = 7.5$.

Our scenario makes several predictions regarding the high-redshift black hole population that distinguish it from the alternative massive seed and super-Eddington accretion models:
\begin{itemize}
    \item chaotic accretion requires the black hole spin to be low; on the other hand, sustained super-Eddington growth requires high spin for AGN jets to remove the generated accretion energy;
    \item if LRDs appear only when energy-driven outflows become capable of removing some gas from the surrounding reservoir, we predict a strong relationship $M \propto \sigma^5$, which should be mostly evident at the lower end of the LRD mass distribution;
    \item the $M-\sigma$ relation should be established at $z \sim 7.5$ and its establishment should coincide with a drastic change in the upper slope of the distribution of maximum SMBH masses as a function of redshift; at higher redshifts, SMBH masses should generally fall under the local $M-\sigma$ relation;
    \item the $M_{\rm BH}/M_*$ ratio should reach a peak at $z \sim 7.5$, because efficiently growing black holes reach accretion rates of order $0.1\times$ the star formation rate in their host galaxies; at later times, the ratio moves steadily closer to the local value.
\end{itemize}

All these predictions will be testable with ongoing and/or upcoming surveys, such as Euclid, Roman, Rubin and NewAthena, that will reveal the mass function of SMBHs at $z > 8$ and the properties of their host galaxies.

\begin{acknowledgements} 
MT and KZ are funded by the Research Council Lithuania grant no. S-MIP-24-100. We thank the anonymous referee for insightful comments that led to a substantial revision of the SMBH growth scenario and the paper structure. 
\end{acknowledgements}


\onecolumn
\appendix

\section{High-z SMBH mass table}

Table \ref{tab:SMBH_data} presents the masses and redshifts of the high-redshift SMBH sample we analysed in this work, ordered by publication date of the earliest mass estimate. To the best of our knowledge, the table contains mass estimates for all currently known SMBHs at $z > 7.5$, in addition to a selection of the most massive SMBHs at $7.0 < z < 7.5$. We divide the sample into four groups based on the mass estimation method used. 

The first is the single direct dynamical mass measurement by \citet{JuodzbalisMaiolino2026} based on resolving the sphere of influence in a strongly lensed galaxy.

The second and largest group contains virial estimates, usually based on the Mg II, H$\beta$ and/or H$\alpha$ line widths. These are our preferred estimates in this work, as they are based directly on observed broad-line properties \citep{VestergaardPeterson2006, Shen2013} and rely on fewer global assumptions about the accretion flow than the alternative methods discussed below. At the same time, they are subject to significant systematic uncertainties. Single-epoch virial estimators inherit the intrinsic scatter of the radius–luminosity relation and depend on the adopted virial scaling, while different calibrations and line recipes can lead to systematic offsets in the inferred masses of $0.3$~dex or more \citep{Shen2013, ShenGrier2024}. Note that some authors choose to exclude this inherent scatter from the uncertainty values of their mass estimates, opting instead to show the formal line fit error and only mentioning the systematic uncertainties in the text. We did not attempt to homogenise the data ourselves. One of the objects in the group, RUBIES-EGS-966323, has three mass estimates based on different ways to fit the galaxy's spectrum with a stellar population template \citep{WangLeja2024}; the highest SMBH mass corresponds to the lowest stellar mass and vice versa.

The third group contains estimates based on UV spectral shape. These estimates are robust up to the Eddington ratio, which must be assumed for all objects. Multiple studies have shown that super-Eddington accretion sometimes occurs, especially in the early Universe \citep{Tang2019MNRAS, Bhatt2024AA, Ighina2025ApJ, Suh2025NatAs}, and beaming of radiation can increase the observed flux above the isotropic value \citep[][also see Section \ref{sec:scenario} above]{King2024MNRAS}, so the true SMBH masses are likely lower than the estimates provided \citep{Lupi2024AA}. A particularly striking example is GN-z11, where the accretion disc spectral fit in the UV region gives an SMBH mass of $M_{\rm BH,UV} \approx 1.1\times10^7\,\msun$ assuming Eddington-limited accretion \citep{Fabian2026MNRAS}, while the virial mass estimate is $M_{\rm BH,vir} \approx 1.6\times10^6\,\msun \approx 0.14 M_{\rm BH,UV}$, which implies super-Eddington accretion with the Eddington ratio $\lambda_{\rm Edd} > 5$ \citep{Maiolino2024AA}. As a result, we treat these mass estimates as less reliable.

The final group contains three estimates and one upper limit based on X-ray luminosity and spectral slope. The dependence on the X-ray spectral slope is evidenced by the two estimates for GHZ9: the higher one is obtained assuming $\Gamma = 2.3$, while the lower one used $\Gamma = 1.8$ \citep{Napolitano2025ApJ}. Furthermore, they depend on the bolometric correction for the conversion between X-ray and bolometric luminosity, which can vary by a factor of three around the mean for high-luminosity quasars \citep{Duras2020AA}. Finally, just as the UV-based estimates, these depend on the assumed Eddington ratio; the quoted values assume it to be equal to unity. So these mass estimates are even less robust than the UV-based ones.

{\tiny

\renewcommand{\arraystretch}{1.3}

\setlength\LTleft{\fill}
\setlength\LTright{\fill}
\setlength{\LTcapwidth}{0.75\textwidth}

\begin{longtable}{lcccc}
\caption{Properties of the black hole sample analysed in this study.} 

\label{tab:SMBH_data}
\\
\hline\hline
Name & $z$ & $\log (M_{\rm BH}/M_\odot)$ & Author & Method \\
\hline &&&& \\[-1.1em]
\endfirsthead

\hline\hline
Name & $z$ & $\log (M_{\rm BH}/M_\odot)$ & Author & Method \\
\hline &&&& \\[-1.1em]
\endhead

\hline
\multicolumn{5}{r}{Continued on next page} \\
\endfoot

\hline\hline
\endlastfoot
Abell2744-QSO1 & 7.04 & $6.94^{+0.15}_{-0.15}$ & \citet{JuodzbalisMaiolino2026} & Direct dynamical mass \vspace{0.2em} \\
\hline &&&& \\[-1.1em]
ULAS J1120+0641 & 7.09 & $9.30^{+0.24}_{-0.19}$ & \citet{MortlockWarren2011} & Broad Mg\,II line (VO09) \\
 &  & $9.38^{+0.10}_{-0.04}$ & \citet{DeRosaVenemans2014} & Broad Mg\,II line (VO09) \\
 &  & $9.04^{+0.01}_{-0.02}$ & \citet{DeRosaVenemans2014} & Broad C\,IV line (VP06) \\
 &  & $9.13^{+0.01}_{-0.01}$ & \citet{YangWang2021} & Broad Mg\,II line (VO09) \\
 &  & $9.28^{+0.40}_{-0.38}$ & \citet{MarshallYue2025} & Broad H$\beta$ line (GH05; VP06) \\
 &  & $9.16^{+0.03}_{-0.04}$ & \citet{BosmanAlvarez-Marquez2025} & Broad H$\alpha$ line (GH05) \\
 &  & $8.95^{+0.08}_{-0.09}$ & \citet{BosmanAlvarez-Marquez2025} & Broad Pa$\alpha$ line (KI10) \\
 &  & $9.07^{+0.06}_{-0.06}$ & \citet{BosmanAlvarez-Marquez2025} & Broad Pa$\beta$ line (KI15) \\
DELS J0038$-$1527 & 7.03 & $9.12^{+0.07}_{-0.09}$ & \citet{WangYang2018} & Broad Mg\,II line (VO09) \\
 &  & $9.08^{+0.11}_{-0.15}$ & \citet{WangYang2018} & Broad Mg\,II line (MLD04) \\
 &  & $9.13^{+0.02}_{-0.02}$ & \citet{YangWang2021} & Broad Mg\,II line (VO09)\\
ULAS J1342+0928 & 7.54 & $8.89^{+0.15}_{-0.12}$ & \citet{BanadosVenemans2018} & Broad Mg\,II line (VO09) \\
 & & $8.91^{+0.09}_{-0.11}$ & \citet{YangWang2021} & Broad Mg\,II line (VO09) \\
 &  & $8.60^{+0.03}_{-0.03}$ & \citet{BosmanAlvarez-Marquez2025} & Broad H$\alpha$ line (GH05) \\
HSC J1243+0100 & 7.07 & $8.52^{+0.21}_{-0.40}$ & \citet{MatsuokaOnoue2019} & Broad Mg\,II line (VO09) \\
DES J0252$-$0503 & 7.00 & $9.14^{+0.05}_{-0.05}$ & \citet{WangDavies2020} & Broad Mg\,II line (VO09) \\
 & & $9.11^{+0.03}_{-0.03}$ & \citet{YangWang2021} & Broad Mg\,II line (VO09) \\
PSO J1007+2115 & 7.52 & $9.18^{+0.05}_{-0.06}$ & \citet{Yang2020ApJ} & Broad Mg\,II line (VO09) \\
\ \ \ \ \ \ \ \ \ (\Pon) & & $9.16^{+0.06}_{-0.07}$ & \citet{YangWang2021} & Broad Mg\,II line (VO09) \\
 &  & $8.85^{+0.02}_{-0.04}$ & \citet{LiuFan2024} & Broad H$\beta$ line (VP06) \\
 & & $9.40^{+0.03}_{-0.04}$ & \citet{LiuFan2024} & Broad H$\beta$ line (VP06) \\
 & & $8.93^{+0.04}_{-0.05}$ & \citet{BosmanAlvarez-Marquez2025} & Broad H$\alpha$ line (GH05) \\
 &  & $8.74^{+0.05}_{-0.06}$ & \citet{BosmanAlvarez-Marquez2025} & Broad Pa$\alpha$ line (KI10) \\
 &  & $8.60^{+0.03}_{-0.03}$ & \citet{BosmanAlvarez-Marquez2025} & Broad Pa$\beta$ line (KI15) \\
J0313$-$1806 & 7.64 & $9.20^{+0.10}_{-0.12}$ & \citet{Wang2021ApJ} & Broad Mg\,II line (VO09) \\
 & & $9.21^{+0.10}_{-0.12}$ & \citet{YangWang2021} & Broad Mg\,II line (VO09) \\
 & & $8.88^{+0.04}_{-0.04}$ & \citet{BosmanAlvarez-Marquez2025} & Broad H$\alpha$ line (GH05) \\
 & & $8.83^{+0.08}_{-0.10}$ & \citet{BosmanAlvarez-Marquez2025} & Broad Pa$\alpha$ line (KI10) \\
 & & $9.02^{+0.05}_{-0.05}$ & \citet{BosmanAlvarez-Marquez2025} & Broad Pa$\beta$ line (KI15) \\
 & & $9.21^{+0.45}_{-0.45}$ & \citet{WolfBanados2026} & Broad H$\beta$ line (SG24) \\
 & & $9.50^{+0.47}_{-0.47}$ & \citet{WolfBanados2026} & Broad Mg\,II line (SG24) \\
CEERS\_1019 & 8.68 & $6.95^{+0.37}_{-0.37}$ & \citet{LarsonFinkelstein2023} & Broad H$\beta$ line (GH05) \\
UNCOVER\_20466 & 8.50 & $8.17^{+0.42}_{-0.42}$ & \citet{KokorevFujimoto2023} & Broad H$\beta$ line (GH05) \\
 & & $8.01^{+0.40}_{-0.40}$ & \citet{KokorevFujimoto2023} & $L_{5100\,\mathrm{\AA}}$ (GH05) \\
 & & $7.45^{+0.40}_{-0.40}$ & \citet{JonesUbler2026} & Broad H$\beta$ line (GH05) \\
 & & $7.57^{+0.40}_{-0.40}$ & \citet{JonesUbler2026} & Broad H$\beta$ line (VP06) \\
 & & $7.58^{+0.40}_{-0.40}$ & \citet{JonesUbler2026} & $L_{5100\,\mathrm{\AA}}$ (VP06) \\  
ZS7 & 7.15 & $7.70^{+0.40}_{-0.40}$ & \citet{UblerMaiolino2024} & Broad H$\beta$ line (GH05) \\
GN-z11 & 10.60 & $6.20^{+0.30}_{-0.30}$ & \citet{Maiolino2024Natur} & Broad Mg\,II line (VO09) \\
RUBIES-EGS-966323 & 8.35 & $6.03^{+0.30}_{-0.28}$ & \citet{WangLeja2024} & Broad H$\beta$ line (AD11) \\
 &  & $8.39^{+0.15}_{-0.12}$ & \citet{WangLeja2024} & Broad H$\beta$ line (AD11) \\
 &  & $8.45^{+0.12}_{-0.14}$ & \citet{WangLeja2024} & Broad H$\beta$ line (AD11) \\
Abell2744-QSO1 & 7.05 & $7.60^{+0.18}_{-0.12}$ & \citet{FurtakLabbe2024} & Broad H$\beta$ line (GH05) \\
 & & $7.59^{+0.30}_{-0.30}$ & \citet{JiMaiolino2025} & Broad H$\alpha$/H$\beta$ line (RV16) \\
COS-66964 & 7.04 & $7.28^{+0.27}_{-0.20}$ & \citet{AkinsCasey2025} & Broad H$\alpha$ line (GH05) \\
CANUCS-LRD-z8.6 & 8.63 & $8.00^{+0.20}_{-0.22}$ & \citet{TripodiMartis2025} & Broad H$\beta$ line (GH05) \\
CAPERS-LRD-z9 & 9.29 & $7.58^{+0.15}_{-0.15}$ & \citet{TaylorKokorev2025} & Broad H$\beta$ line (GH05) \\
J1007\_AGN & 7.26 & $8.65^{+0.33}_{-0.31}$ & \citet{SchindlerHennawi2025} & Broad H$\beta$ line (GH05) \\
JADES GS-20057765 & 8.91 & $7.33^{+0.62}_{-0.70}$ & \citet{JuodzbalisMaiolino2026} & Broad H$\alpha$ line (RV16) \\
JADES GN-4685 & 7.42 & $7.36^{+0.45}_{-0.42}$ & \citet{JuodzbalisMaiolino2026} & Broad H$\alpha$ line (RV16) \\
JADES GS-164055 & 7.40 & $7.63^{+0.59}_{-0.66}$ & \citet{JuodzbalisMaiolino2026} & Broad H$\alpha$ line (RV16) \\
JADES GS-20030333 & 7.89 & $7.42^{+0.65}_{-0.48}$ & \citet{JuodzbalisMaiolino2026} & Broad H$\alpha$ line (RV16) \\
GNz7q & 7.19 & $7.55^{+0.34}_{-0.34}$ & \citet{FeiFujimoto2026} & Broad H$\beta$ line (DH15) \\
C3D-z7AGN-1 & 7.65 & $8.60^{+0.35}_{-0.33}$ & \citet{LinFan2026} & Broad H$\beta$ line (GH05) \vspace{0.2em} \\
EUCL J052209.82-512709.2 & 7.50 & $7.60^{+0.4}_{-0.4}$ & \citet{Yang2026AA} & Broad C IV line (VP06) \vspace{0.2em} \\
\hline &&&& \\[-1.1em]
MoM-z14 & 14.44 & $6.90$ & \citet{Fabian2026MNRAS} & Accretion disc fit \\
JADES GS-z14-1 & 13.86 & $6.40$ & \citet{Fabian2026MNRAS} & Accretion disc fit \\
GHZ2 (GLASS-z12) & 12.34 & $6.90$ & \citet{Fabian2026MNRAS} & Accretion disc fit \\
 &  & $7.20^{+0.04}_{-0.04}$ & \citet{ChavezOrtizFinkelstein2025} & SED modelling \\
GN-z11 & 10.60 & $7.05$ & \citet{Fabian2026MNRAS} & Accretion disc fit \\
COSW-106725 & 7.70 & $\geq 8.81$ & \citet{LambridesChiaberge2024} & SED modelling \\
ULAS J1342+0928 & 7.54 & $9.20^{+0.20}_{-0.20}$ & \cite{TrefoloniCarniani2025} & Accretion disc fit \vspace{0.2em} \\
\hline &&&& \\[-1.1em]
UHZ1 & 10.30 & $7.60^{+0.30}_{-0.30}$ & \citet{BogdanGoulding2024} & X-ray luminosity \\
UNCOVER DR1 13556 & 8.50 & $<7.20$ & \citet{AnannaBogdan2024} & X-ray luminosity upper limit \\
GHZ9 & 10.14 & $8.20^{+0.08}_{-0.09}$ & \citet{Napolitano2025ApJ} & X-ray luminosity ($\Gamma = 2.3$) \\
 & & $7.68^{+0.07}_{-0.09}$ & \citet{Napolitano2025ApJ} & X-ray luminosity ($\Gamma = 1.8$) \vspace{0.2em} \\
\end{longtable}
\tablefoot{The columns show, from left to right, the object name, its redshift, SMBH mass estimate (with uncertainties where available), reference for the mass estimate and the method used to estimate the mass. Horizontal lines separate the sample into groups based on mass estimation method: direct dynamical mass, virial, accretion-disc or SED modelling and X-ray luminosity, respectively. Note that we include the method column as a broad categorisation and to allow for easier differentiation between values. We abbreviate the often used calibrations as follows: MLD04: \citet[][]{McLureDunlop2004}, GH05: \citet[][]{GreeneHo2005}, VP06: \citet[][]{VestergaardPeterson2006}, VO09: \citet[][]{VestergaardOsmer2009}, KI10: \citet[][]{KimIm2010}, AD11: \citet[][]{AssefDenney2011}, KI15: \citet[][]{KimIm2015}, DH15: \citet[][]{DuHu2015}, RV16: \citet[][]{ReinesVolonteri2015}, SG24: \citet[][]{ShenGrier2024}. In cases where no method reference is given, the method is described primarily in the mass estimate paper.}

}

\label{lastpage}
\end{document}